\documentclass[sigconf, nonacm, pdfa]{acmart}

\usepackage[a-2b]{pdfx}
\usepackage{amsthm}
\usepackage{amsmath,amsfonts}
\usepackage[linesnumbered,ruled,vlined]{algorithm2e}
\usepackage{balance}
\usepackage{multicol}
\usepackage{array}
\usepackage{booktabs}
\usepackage[nameinlink]{cleveref}
\usepackage{seqsplit}
\usepackage[subrefformat=parens]{subcaption}
\usepackage{textcomp}
\usepackage{stfloats}
\usepackage{url}
\usepackage{verbatim}
\usepackage{graphicx}
\usepackage{xcolor}

\newcommand{\adjustCaption}{\vspace{-0.0em}}

\SetCommentSty{mycommfont}
\SetKwProg{Function}{Function}{}{}

\theoremstyle{plain}

\newtheorem*{theorem*}{Theorem}

\newcommand\vldbpagestyle{plain} 

\begin{document}
\title{Oze: Decentralized Graph-based Concurrency Control for Long-running Update Transactions (Extended Version)}

\settopmatter{authorsperrow=4}

\author{Jun Nemoto}
\affiliation{%
  \institution{Scalar, Inc.}
}
\email{jun@scalar-labs.com}

\author{Takashi Kambayashi}
\affiliation{%
  \institution{Nautilus Technologies, Inc.}
}
\email{kambayashi@nautilus-technologies.com}

\author{Takashi Hoshino}
\affiliation{%
  \institution{Cybozu Labs, Inc.}
}
\email{hoshino@labs.cybozu.co.jp}

\author{Hideyuki Kawashima}
\affiliation{%
  \institution{Keio University}
}
\email{river@sfc.keio.ac.jp}

\begin{abstract}
This paper proposes Oze, a concurrency control protocol that handles heterogeneous workloads, including long-running update transactions. Oze explores a large scheduling space using a multi-version serialization graph to reduce false positives. Oze manages the graph in a decentralized manner to exploit many cores in modern servers. We further propose an OLTP benchmark, BoMB (Bill of Materials Benchmark), based on a use case in an actual manufacturing company. BoMB consists of one long-running update transaction and five short transactions that conflict with each other. Experiments using BoMB show that Oze can handle the long-running update transaction while achieving four orders of magnitude higher throughput than state-of-the-art optimistic and multi-version protocols and up to five times higher throughput than pessimistic protocols. We also show Oze performs comparably with existing techniques even in a typical OLTP workload, TPC-C, thanks to a protocol switching mechanism.
\end{abstract}

\maketitle

\pagestyle{\vldbpagestyle}

\section{Introduction} \label{sec:intro}

\subsection{Motivation}

Conventional serializable concurrency control protocols are not well-suited for certain real-world OLTP workloads, especially those involving long-running update transactions. A typical example can be found in product costing systems used in manufacturing industries. These systems perform long-running update transactions based on a bill of materials (BoM), and such transactions must be isolated from others in a serializable manner to ensure accurate and optimal resource planning.

To isolate the long transaction from other transactions,
traditional systems usually run it where short online transactions are absent, e.g., at night~\cite{Bog14} or use stale materialized views~\cite{Müller13}.
However, this approach is becoming impractical. Modern product costing must handle frequent updates to materials and their costs, especially in the face of supply chain disruptions. This trend necessitates on-demand costing, which requires the ability to execute long and short transactions concurrently while preserving serializability.
Snapshot isolation~\cite{Berenson95} cannot guarantee the consistency of BoM trees used by the long transactions for product costing, and it produces incorrect results, as discussed in~\Cref{sec:bomb-transaction}.

\begin{figure}[t]
  \centering
  \includegraphics[keepaspectratio, scale=0.40]{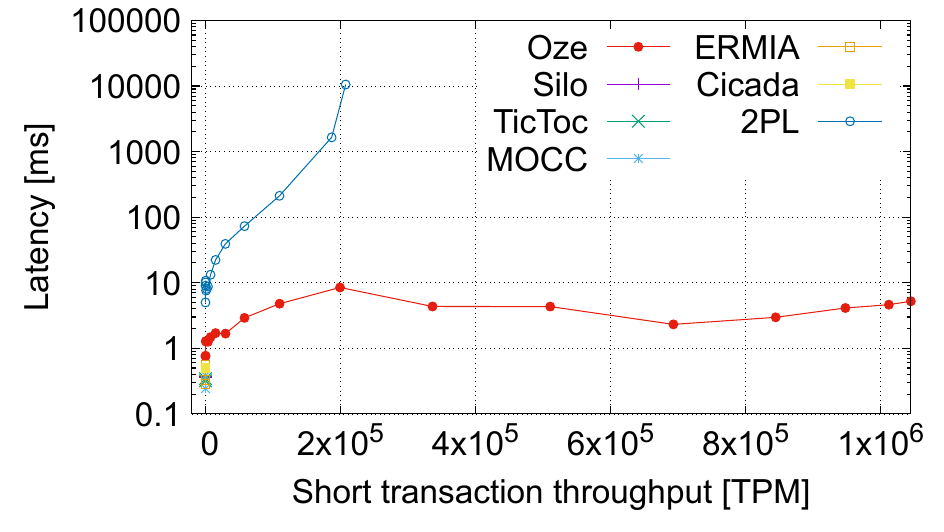}
  \adjustCaption
  \caption{Throughput versus average latency of short transactions in BoM benchmark. Existing OCC and MVCC protocols cannot increase throughput any further because they will not be able to commit the long transaction (L1). 2PL shows lower throughput with higher latency than Oze because it has to wait for the L1 commit for a long time.}
  \label{fig:cc-comparison-tps-vs-latency}
\end{figure}

\subsection{Challenges}

None of the conventional protocols~\cite{Tu13,Wang16,Yu16,Wang17,Lim17} can handle the new workload with BoM. As shown in \Cref{fig:cc-comparison-tps-vs-latency}, the throughput and latency of the short transactions degrade when attempting to ensure the successful commit of a long transaction (referred to as L1) in our BoM-based benchmark.
Existing OCC and MVCC protocols show lower throughput because they are more likely to abort the L1 transaction as the request rate increases.
All of the L1 aborts in those protocols are false positives (i.e., false aborts), which result from protocol-specific constraints on transaction ordering.
2PL exhibits better throughput than OCC and MVCC; however, the L1 transactions significantly reduce the throughput of short transactions and increase the latency, as shown in the figure, because they must wait for a lock held by the L1 transaction for a long time. See \Cref{sec:soa-protocols} for details of these behaviors in each protocol.

Deterministic approaches~\cite{Thomson12,Fan19} can handle this workload under a certain condition, where BoMs do not change before and after the L1 transaction. We call it a static BoM. However, our target BoM is often updated to dynamically reflect the effect of supply chain disruption, and on-demand costing is required. We call such a BoM a dynamic BoM. Despite using reconnaissance queries~\cite{Thomson12} to identify BoM trees beforehand, deterministic approaches struggle to commit the L1 transaction without sacrificing the short transaction throughput in a dynamic BoM. Our deterministic extension of 2PL, called D2PL, showed only marginal performance, as depicted in~\Cref{fig:dbom-cc-comparison} in~\Cref{sec:eval-bomb}.

It is challenging for a serializable protocol to ensure committing long transactions without degrading the performance of short transactions running concurrently.

\subsection{Contributions and paper organization}

\textbf{Oze Protocol: } We propose Oze, a concurrency control protocol that can handle long-running update transactions while achieving high short transaction throughput with reasonable latency, as shown in Figure~\ref{fig:cc-comparison-tps-vs-latency}. The key ideas of Oze are twofold.

The first idea is to exploit the large scheduling space using a multi-version serialization graph (MVSG)~\cite{Bernstein83} in a decentralized manner. Conventional MVSG-based protocols such as MVSGT~\cite{Hadzilacos85} and MVSGA~\cite{Hadzilacos88} assume a single centralized graph management and thus fail to exploit modern many-core architectures.
Oze manages a record-local graph associated with each record, each of which represents only the dependencies among transactions accessing that specific record, and it partially synchronizes these graphs across records to check for acyclicity.

The second idea is dynamic protocol switching. To handle conventional workloads such as TPC-C~\cite{tpcc}, Oze provides the OCC mode in addition to the MVSG mode and switches them bidirectionally and dynamically depending on the workloads, as depicted in~\Cref{fig:protocol-switch} and in~\Cref{fig:tpcc}.

\textbf{BoMB Benchmark: } We also present an OLTP benchmark, BoMB (Bill of Materials Benchmark), which emulates real-world on-demand product costing. Conventional OLTP benchmarks do not contain long transactions with consecutive anti-dependencies. TPC-C~\cite{tpcc} and TPC-E~\cite{tpce} do not have any long transactions. TPC-EH~\cite{Wang17} contains a long-running transaction that includes write operations, but the result of writes is never read by other transactions, and thus it does not produce consecutive anti-dependencies. HTAP benchmarks like OLxPBench~\cite{Kang23} contain long \textit{read-only} transactions.
BoMB includes a long-running update transaction that will be falsely aborted with high probability in conventional serializable protocols due to two consecutive and long-term anti-dependency edges created by the other five short transactions. We design BoMB based on a workload in a real bread manufacturing company, Andersen~\cite{Andersen}, with diverse configurable parameters, allowing BoMB to represent a variety of BoM workloads.

We evaluate Oze with modern concurrency control protocols using BoMB on CCBench~\cite{Tanabe20}. Experimental results show that only Oze can ensure the L1 transaction commit without sacrificing the short transaction throughput.

The rest of this paper is organized as follows. First, in Section~\ref{sec:bomb}, we describe the workload in BoMB and why it is challenging for existing protocols. Next, we describe the design and implementation of the Oze protocol in Section~\ref{sec:design} and Section~\ref{sec:implementation}, respectively. In Section~\ref{sec:eval}, we evaluate several protocols using BoMB and TPC-C. In Section~\ref{sec:related}, we describe related work. Finally, we conclude this paper in Section~\ref{sec:conclusion}.

\section{BoMB and Challenges} \label{sec:bomb}
This section presents a new OLTP benchmark, BoMB (BoM Benchmark). First, we provide an overview of BoMB's database and workload. We then show how and why existing protocols cannot effectively handle the BoMB workload.

\subsection{Overview}

\begin{figure}
    \centering
    \includegraphics[keepaspectratio, scale=0.5]{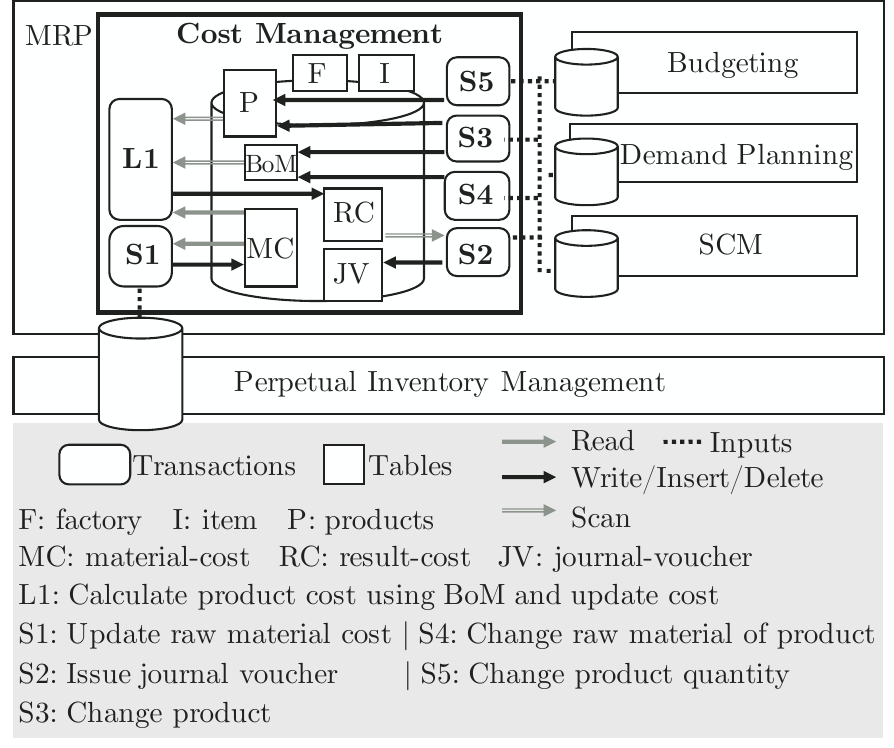}
    \adjustCaption
    \caption{System and workload overview.}
    \label{fig:workload}
\end{figure}

We model on-demand product costing based on a real bread manufacturing company, Andersen~\cite{Andersen}, extracting common components and parameters applicable to various industries. Currently, they completely isolate the product costing from the OLTP database. They export input data from the OLTP database to a Hadoop-based distributed system, calculate product costs, and write back the results into the OLTP database. To avoid any kind of inconsistency, they do not allow reading and writing the OLTP database during the cost calculation. Therefore, serializable isolation is required for on-demand costing as well.

\Cref{fig:workload} shows the modeled system that
includes a manufacturing resource planning (MRP) system and an inventory management system, with modules for cost management, budgeting, demand planning, and supply chain management (SCM). The cost management module, involving long-running update transactions, produces the most complex and challenging workloads. Our model focuses on this module, identifying one long transaction (L1) and five major short transactions (S1-S5) that are common in manufacturing. L and S stand for "long" and "short," respectively. All these transactions commonly occur in manufacturing industries~\cite{Younus10,OpenBOM} and can be widely applied beyond the bread company.

\subsection{Tables} \label{sec:table-and-params}

\begin{table}[t]
    \caption{BoMB parameters.}
    \adjustCaption
    \label{tab:params}
    \begin{tabular}{lcl}
    \toprule

    \begin{tabular}{l}
        Parameters
    \end{tabular} &
        Default &
    \begin{tabular}{l}
        Description
    \end{tabular} \\

    \midrule

    \begin{tabular}{l}
        factories
    \end{tabular} &
        8 &
    \begin{tabular}{l}
        \# of factories
    \end{tabular} \\

    \begin{tabular}{l}
        product-types
    \end{tabular} &
        72,000 &
    \begin{tabular}{l}
        \# of product types
    \end{tabular} \\

    \begin{tabular}{l}
        material-types
    \end{tabular} &
        198,000 &
    \begin{tabular}{l}
        \# of material types
    \end{tabular} \\

    \begin{tabular}{l}
        raw-material-types
    \end{tabular} &
        75,000 &
    \begin{tabular}{p{0.35\linewidth}}
        \# of raw material types
    \end{tabular} \\

    \begin{tabular}{l}
        material-trees-\\per-product
    \end{tabular} &
        5 &
    \begin{tabular}{l}
        \# of material trees \\ per product
    \end{tabular} \\
  
    \begin{tabular}{l}
        material-tree-size
    \end{tabular} &
        10 &
    \begin{tabular}{l}
        \# of materials \\ in a material tree
    \end{tabular} \\
    
    \begin{tabular}{l}
        raw-materials-\\per-leaf
    \end{tabular} &
        3 &
    \begin{tabular}{p{0.35\linewidth}}
        \# of raw materials in a leaf material
    \end{tabular} \\

    \begin{tabular}{l}
        target-products
    \end{tabular} &
        100 &
    \begin{tabular}{p{0.35\linewidth}}
        \# of products manufactured in a factory
    \end{tabular} \\

    \begin{tabular}{l}
        target-materials
    \end{tabular} &
        1 &
    \begin{tabular}{p{0.35\linewidth}}
        \# of raw materials \\ for update
    \end{tabular} \\

    \bottomrule
    \end{tabular}
\end{table}

BoMB uses the following seven tables. The underlined attribute is the primary key. Note that \texttt{INT16}, \texttt{INT32}, and \texttt{INT64} are integers of 16, 32, and 64 bits, respectively. Adjustable parameters, including cardinalities for the BoMB, are shown in Table~\ref{tab:params}.

\texttt{\textbf{factory}(\underline{id} INT32, name VARCHAR)}: We assume multiple factories in a company. The \texttt{factory} table manages a list of those factories. The number of factories is set by the parameter \texttt {factories}.

\texttt{\textbf{item}(\underline{id} INT32, name VARCHAR, type INT16)}: The \texttt{item} table manages the name of items with their type: product, material, or raw material. The \texttt{item} table stores the total records of products (\texttt{product-types}), materials (\texttt{\seqsplit{material-types}}), and raw materials (\texttt{\seqsplit{raw-material-types}}).

\texttt{\textbf{product}(\underline{factory\_id} INT32, \underline{item\_id} INT32, quantity DOUBLE)}: 
The \texttt{product} table manages the manufactured products and their quantity in each factory. In product costing, it is used to obtain the products currently in production at the factory.

\texttt{\textbf{bom}(\underline{parent\_item\_id} INT32, \underline{child\_item\_id} INT32, quan\-ti\-ty DOUBLE)}: The \texttt{bom} table manages a list of (intermediate and raw) materials and the quantities of each needed to manufacture a product. Specifically, it stores the parent item ID, child item ID, and quantity in each record and hierarchically represents BoM trees. Details of the structure of BoM trees and product costing using this table are described in Section~\ref{sec:bom-tree}.

\texttt{\textbf{material-cost}(\underline{factory\_id} INT32, \underline{item\_id} INT32, \seqsplit{stock\_quantity} DOUBLE, stock\_amount DOUBLE)}: The \texttt{\seqsplit{material-cost}} table manages the costs of raw materials, the stock quantity, and the amount of the raw materials for each factory and item.

\texttt{\textbf{result-cost}(\underline{factory\_id} INT32, \underline{item\_id} INT32, cost DOUBLE)}: The \texttt{result-cost} table contains the latest cost calculation results for each product in each factory.

\texttt{\textbf{journal-voucher}(\underline{voucher\_id} INT64, date DATE, debit INT32, credit INT32, amount DOUBLE, description VARCHAR)}: The \texttt{journal-voucher} table manages journal vouchers, which are issued when manufacturing processes are executed using the calculated costs and utilized by each related module, e.g., budgeting, demand planning, and SCM.

\subsection{BoM Tree} \label{sec:bom-tree}

\begin{figure}[t]
    \centering
    \includegraphics[keepaspectratio, scale=0.52]{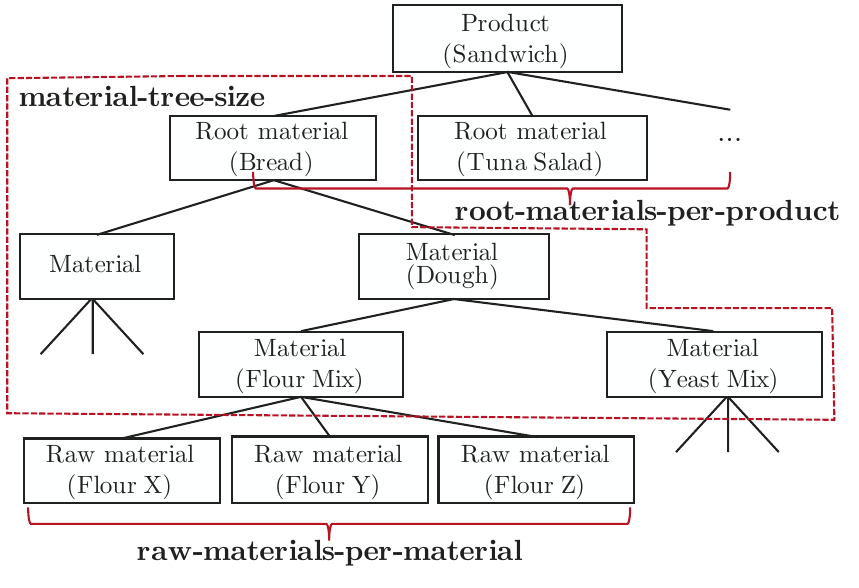}
    \adjustCaption
    \caption{Example of BoM tree.}
    \label{fig:bom-tree}
\end{figure}

The \texttt{bom} table represents an item along with its constituent child items and their quantities. Such a relationship can be logically expressed in a tree structure, which we call a BoM tree. A record in the \texttt{bom} table is identified by a parent item ID and the child item ID. To obtain the components of an item, the \texttt{bom} table is scanned by specifying the item ID as a parent. An example of a BoM tree is shown in Figure~\ref{fig:bom-tree}. The product consists of several major materials (hereafter, "root materials" for convenience). For example, in the production of sandwiches, the root materials correspond to bread and the ingredients inside (e.g., tuna salad). Each root material is made from multiple materials. For example, for bread, the material is dough, and the raw materials are flour, yeast, etc.

When starting the benchmark, the BoM trees are initialized as follows. (1) Select a set of materials with size \texttt{\seqsplit{material\_tree\_size}}. (2) Select the root material from them. (3) Add the remaining materials as child nodes to random tree nodes. (4) Add \texttt{\seqsplit{raw-materials-per-leaf}} raw materials to each leaf of the tree. Raw materials are randomly selected from \texttt{\seqsplit{raw-materials-types}}. (5) After generating all trees until \texttt{\seqsplit{materials-types}} materials are exhausted, assign \texttt{\seqsplit{material-trees-per-product}} trees to each product. Although the tree structure is randomly configured by default for versatility in BoMB, skew may be given depending on the target industry.

The product cost is calculated using the BoM tree as follows. (1) Identify the products to be costed. (2) Construct a BoM tree by referring to the \texttt{bom} table and recursively acquiring the materials that comprise each product. Each tree node contains an item ID, a list of child item IDs, a unit price, and a required quantity. (3) Set the unit cost for each raw material, which is a leaf node of the BoM tree, by referring to the \texttt{\seqsplit{material-cost}} table. (4) Calculate the product cost by recursively calling the \texttt{calculate\_cost()} function shown in~\Cref{alg:cost} from the root node of the BoM tree.

\begin{algorithm}[t]
  \small
  \caption{Calculate cost of product}\label{alg:cost}
  \Function{calculate\_cost()}{
      \If{is\_leaf()}{
          \Return{unit\_cost*quantity}
      }
      subtotal = 0 \\
      \For{child \textbf{in} children}{
          subtotal += child->calculate\_cost()
      }
      \Return{subtotal*quantity}
  }
\end{algorithm}

\subsection{Transactions} \label{sec:bomb-transaction}

\textbf{Long transaction: }
The L1 (update-product-cost) transaction is a long transaction that builds BoM trees described in Section~\ref{sec:bom-tree} and calculates product costs. First, it selects one factory randomly and obtains all products manufactured at the factory and their \texttt{quantity} by referring to the \texttt{product} table. Next, it builds a BoM tree for the product and calculates the cost. When building the BoM tree, it refers to the \texttt{material-cost} table, updated by S1 transactions, in addition to the \texttt{bom} table. Then, it writes the result to the \texttt{result-cost} table, which is referred to by S2 transactions. These steps are repeated for all products. All products must be handled in a single transaction because optimal production planning must be performed based on a consistent view of the costing results throughout a factory. Since S2 transactions scan all product costs in a factory to guarantee such a consistent view, transaction chopping~\cite{Shasha95} cannot be applied without violating serializability. We use a parameter, \texttt{\seqsplit {target-products}} that represents how many products are currently manufactured at each factory (default: 100). This parameter is for adjusting the length of the transaction. When an L1 transaction is executed with the default setting, it reads/scans about 20,000 records in total for calculating costs and writes 100 records as the result.

\textbf{Short transactions: }
The S1 (update-material-cost) transaction is a short transaction that changes the cost of raw materials. It performs read-modify-write operations on randomly selected records in the \texttt{material-cost} table. The S2 (issue-journal-voucher) transaction is a short transaction that creates a journal voucher based on the calculated product cost. It scans the \texttt{\seqsplit{result-cost}} table to obtain the costs of all products in a factory and inserts the voucher records into the \texttt{journal-voucher} table. The S3 (change-product) transaction is a short transaction that replaces an old product with a newly-developed product. It creates a new BoM tree for the product, inserts the records into the \texttt{bom} table, and replaces the product by deleting and inserting the record from/into the \texttt{product} table. The S4 (change-raw-material) transaction is a short transaction that replaces a raw material of a product with a different one (e.g., change a flour X to X'). It deletes and inserts a raw material record for a BoM tree in the \texttt{bom} table. The S5 (change-product quantity) transaction is a short transaction that updates a manufacturing quantity of a product in a factory. It updates a record in the \texttt{product} table.

\textbf{Transaction mix: }
BoMB can run with two settings based on the characteristics of the target BoM: \textit{static} and \textit{dynamic} BoM. For static BoM, BoMB only runs L1, S1, and S2 transactions. For dynamic BoM, it additionally runs S3, S4, and S5 transactions, which change BoM structures dynamically. We set the default percentage of short transactions for static BoM to be 50\% each for S1 and S2. Similarly, for dynamic BoM, we set the ratio at 45\%, 45\%, 1\%, 1\%, and 8\% for S1, S2, S3, S4, and S5, respectively. This default transaction mix is configured based on a case for our industrial collaborators. Since it may vary depending on industries and companies, BoMB allows users to customize the transaction mix to meet their requirements.

\textbf{Regulation: }
According to the discussion with our industrial collaborators, for the BoMB workload, the mandatory requirement is to ensure the commit of L1. Even if a protocol can accidentally commit L1 by repeating aborts and retries for a long time, this can delay the cost calculations and result in incorrect product pricing and profit calculations, affecting business decisions. To reflect this requirement, we define the BoMB benchmark score as the maximum short transaction throughput that can be achieved with less than 1\% abort rate of L1\footnote{We choose 1\% as a realistic compromise since it is theoretically impossible for protocols such as OCC to guarantee 0\% abort rate under the condition where all transactions run concurrently.}. To always be running L1 transactions and short transactions concurrently, a new L1 transaction must start just after the previous L1 commit when measuring the short transaction throughput. Moreover, to ensure accurate product costing, all transactions must be executed with serializable isolation.

\subsection{Why serializable?}

Product costing with lower isolation levels, e.g., snapshot isolation~\cite{Berenson95}, easily causes anomalies. Suppose transactions that change raw materials, such as S4, and they must be executed with a certain condition regarding other intermediate materials; e.g., if a flour mix X consists of flour A and B, then yeast P in a yeast mix Y should be replaced with yeast P' or if a flour mix X consists of flour A and B', then yeast Q in Y should be replaced with yeast Q'. Such a transaction and another transaction that replaces B with B' may cause anomalies for L1 transactions, i.e., L1 observes an inconsistent combination of those materials as a BoM tree. This can happen even if all transactions except for L1 run with serializable isolation. The detailed situation is as follows:

\begin{itemize}
  \item Intermediate material X consists of \{A, B\}.
  \item Intermediate material Y consists of \{P, Q\}.
  \item T1 changes a raw material from B to B'.
  \item T2 changes a raw material from P to P' if X consists of \{A, B\} or from Q to Q' if X consists of \{A, B'\}.
  \item T3 scans X and Y and updates a different table like the L1 transaction.
\end{itemize}

Then, the following history can happen under snapshot isolation; each operation, s, i, d, and c, shows scan, insert, delete, and commit, respectively. The write operations of T3 are omitted.

\begin{equation*}
  \begin{aligned}
  h = s_2(X, \{A, B\}) \ s_2(Y, \{P, Q\}) \ s_1(X, \{A, B\}) \ d_1(B) \ i_1(B') \ c_1 \\
  s_3(X, \{A, B'\}) \ s_3(Y, \{P, Q\}) \ d_2(P) \ i_2(P') \ c_2 \ c_3
  \end{aligned}
\end{equation*}

This is an anomaly that T3 observes \{A, B'\} and \{P, Q\} even though the final result shows \{A, B'\} and \{P', Q\}. This cannot happen in any serializable execution since if B was changed to be B' before changing a raw material of Y, the final result should be \{A, B'\} \{P, Q'\}. In other words, the combination \{A, B'\} and \{P, Q\} observed by T3 is a BoM tree that cannot exist. Although T3 is not a read-only transaction, this anomaly has the same structure as a typical read-only anomaly described in the literature~\cite{Fekete04}.

Even a slight change in the short transactions of BoMB can easily cause such a problem with the correctness of L1. Under lower isolation levels, verifying the correctness of schedules or their safety in application semantics requires significant effort. Thus, executing all transactions, including L1, with serializable isolation is the most practical solution to avoid incorrect results here.

\begin{figure*}
    \begin{minipage}{0.195\linewidth}
      \centering
      \includegraphics[keepaspectratio, scale=0.38]{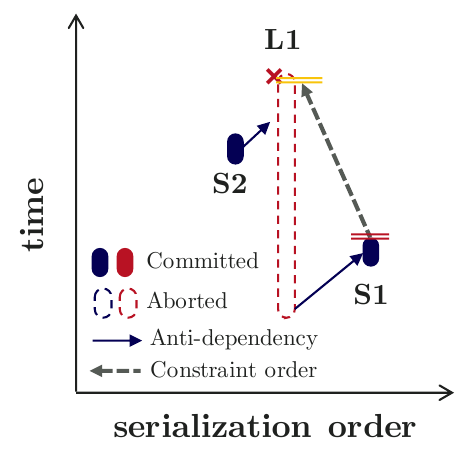}
      \subcaption{OCC~\cite{Tu13,Yu16}}
    \end{minipage}
    \begin{minipage}{0.195\linewidth}
      \centering
      \includegraphics[keepaspectratio, scale=0.38]{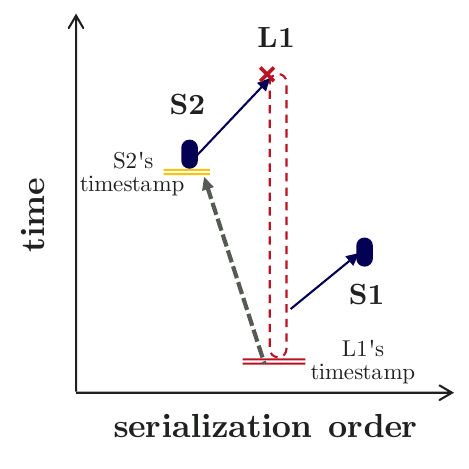}
      \subcaption{MVTO~\cite{Lim17}}
    \end{minipage} 
    \begin{minipage}{0.195\linewidth}
      \centering
      \includegraphics[keepaspectratio, scale=0.38]{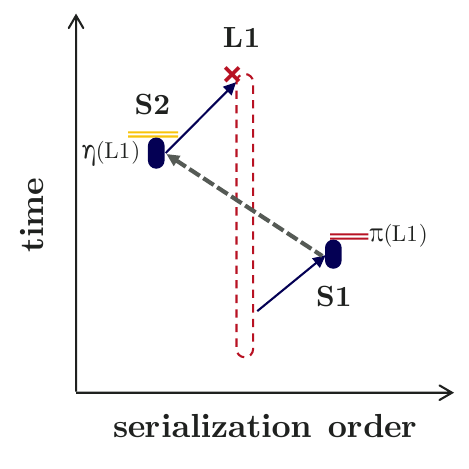}
      \subcaption{SSN~\cite{Wang17}}
    \end{minipage}
    \begin{minipage}{0.195\linewidth}
      \centering
      \includegraphics[keepaspectratio, scale=0.38]{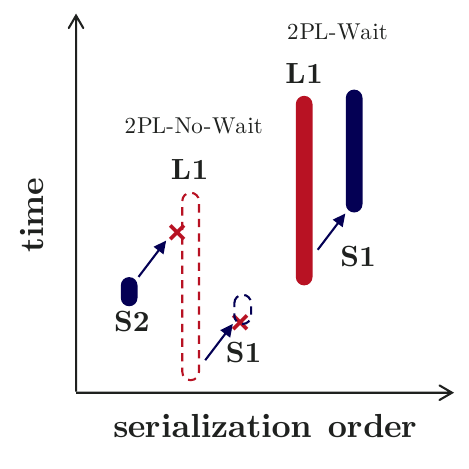}
      \subcaption{2PL~\cite{Bernstein81}}
    \end{minipage}
    \begin{minipage}{0.195\linewidth}
      \centering
      \includegraphics[keepaspectratio, scale=0.38]{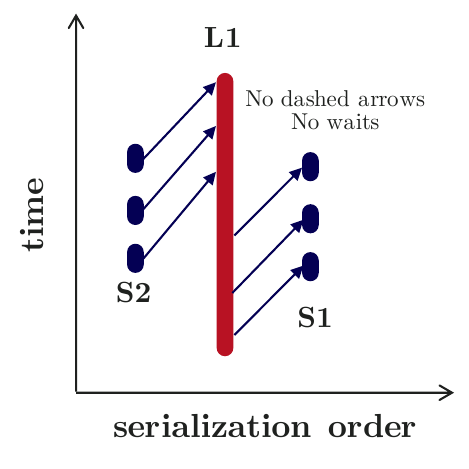}
      \subcaption{Oze}
    \end{minipage}
    \adjustCaption
    \caption{Why existing protocols are not enough? (a)--(c) L1 aborts with false positives because the observed anti-dependency order (blue arrows) does not match the constraint order (dashed arrows, see~\Cref{sec:soa-protocols} for details): (a) commit order, (b) begin timestamp order, and (c) commit-timestamp-based order. (d) L1 aborts, or S1 must wait for a long time. (e) Oze can commit L1 without false positives or S1's waits, thanks to precise order tracking.}
    \label{fig:cc-analysis}
\end{figure*}

\subsection{BoMB with Conventional Protocols and Challenges} \label{sec:soa-protocols}

\Cref{fig:cc-analysis} illustrates why existing concurrency control protocols fail to handle BoMB transactions. Inherently, transaction serializability can be determined solely by the dependency order, i.e., reads-from relation and version order, including anti-dependencies decided by version order~\cite{Weikum01}. To reduce the cost of tracking all of those dependencies, conventional protocols impose a certain constraint on the acceptable serialization order. This protocol-specific constraint is an order, and we call it constraint order. If the observed dependency order (including potential ones) of a transaction does not match the constraint order, the transaction is aborted. Most of the existing protocols abort the L1 transaction because the dependency order of BoMB transactions (S2$<$L1$<$S1) does not obey their own constraint orders, although this is a false positive.

For example, OCC protocols such as Silo~\cite{Tu13} and TicToc~\cite{Yu16} force transactions to obey a constraint order, which is an order in which they entered the serialization point by acquiring the necessary locks in the validation phase. The OCC-specific behavior, such as aborts due to finding an overwriter, can be explained as a mismatch between a potential anti-dependency order (L1$<$S1) and the constraint order (S1$<$L1), as shown in~\Cref{fig:cc-analysis}(a). Due to the constraint order, the L1 transaction is aborted as a false positive even though they are serializable (S2$<$L1$<$S1).

MOCC~\cite{Wang16} combines OCC with a pessimistic scheme using locks and a hotspot counter. Since not all records are treated pessimistically, the L1 transaction still fails read validation. Timestamp adjustments used in some OCC variants~\cite{Boksenbaum87,Lee93,Kwok96} do not contribute to the L1 completion with or without priority setting because the room for the adjustment is exhausted in a moment by several S1 transactions around the L1 transaction.

MVTO forces transactions to obey the begin timestamp order as the constraint order. Because the observed anti-dependency (S2$<$L1) of an L1 transaction is opposite to the begin timestamp order (L1$<$S2), the L1 transaction is aborted, as shown in~\Cref{fig:cc-analysis}(b).

Serial safety net (SSN)~\cite{Wang17} in ERMIA~\cite{Kim16} uses a \seqsplit{commit-timestamp-based} order as the constraint order. This order is adjusted from the actual commit order based on the two watermarks of a transaction $T$: $\pi(T)$ and $\eta(T)$. Because the anti-dependency order (S2$<$L1$<$S1) observed by the L1 transaction using $\pi(L1)$ and $\eta(L1)$ is inconsistent with the commit-timestamp-based order (S1$<$S2), the L1 transaction is aborted, as shown in~\Cref{fig:cc-analysis}(c).

2PL and its variants~\cite{Rosenkrantz78,Bernstein81,Weikum01,Corbett13,Guo21} are protocols that serialize transactions by using the locking order as the constraint order. As shown in~\Cref{fig:cc-analysis}(d), in 2PL-based protocols, conflicting transactions cannot run concurrently because they need to obey the constraint order by waiting for locks. Specifically, in 2PL-Wait~\cite{Bernstein81}, S1 transactions must keep waiting for the L1 completion. Similarly, in 2PL-No-Wait~\cite{Bernstein81}, S1 transactions must stall because they continue to abort and retry. In addition, L1 transactions may abort when they access records already locked by S1 or S2 transactions.

\textbf{Challenges: }
As stated above, OCC and MVCC protocols force a constraint order in addition to the dependency order. While these protocols can efficiently check serializability, they limit the scheduling space to be explored by the constraint order and suffer from false aborts. 2PL-based approaches can commit L1 transactions since 2PL forces transactions to wait so that the transactions obey the constraint order decided by the locking order. However, this locking order limits the scheduling space and significantly reduces the short transaction performance if the preceding L1 transactions exist. Therefore, it is challenging to achieve no false aborts for the long transactions without reducing performance for the short transactions running in parallel.

\section{Oze Design} \label{sec:design}

To avoid false positives and performance degradation due to the constraint order, we propose Oze, which is an MVSG-based protocol and does not depend on such a constraint at all. Oze also supports dynamic protocol switching between the MVSG mode and the OCC mode to efficiently handle conventional workloads.

\subsection{Decentralizing MVSG} \label{sec:decentralizing-mvsg}

The literature~\cite{Bernstein83} proposed the multiversion serialization graph (MVSG) and proved that a transaction schedule is serializable iff there exists an acyclic MVSG. An MVSG is formed by a multiversion schedule $s$ and a version order $\ll$. A multiversion schedule contains two types of operations: a write operation $w_i(x_i)$ that denotes a transaction $t_i$ writes a version of a record $x$ and a read operation $r_j(x_i)$ that denotes $t_j$ reads $x_i$ written by $t_i$, which is called a reads-from relation. A version order $\ll$ for $s$ is the union of all version orders of data items written by operations in $s$, where a version order for $x$ is any non-reflexive and total ordering of all versions of $x$ and $x_i \ll x_j$ denotes that a version $x_i$ precedes another version $x_j$. For given $s$ and $\ll$, $MVSG(s, \ll)$ has nodes for a set of transactions $t_0 \dots t_N$ and edges as follows: for distinct operations $w_j(x_j)$, $r_i(x_j)$, and $w_k(x_k)$ in $s$ where $t_i \neq t_k$,

\begin{itemize}
  \item $t_j \rightarrow t_i$ for a reads-from relation $r_i(x_j)$ (\textit{wr}-dependency)
  \item $t_i \rightarrow t_k$ if $x_j \ll x_k$ (\textit{rw}-dependency)
  \item $t_k \rightarrow t_j$ if $x_k \ll x_j$ (\textit{ww}-dependency)
\end{itemize}

Conventional MVSG-based protocols~\cite{Hadzilacos85,Hadzilacos88} are not scalable because they assume a single centralized graph structure for the MVSG. A giant lock is necessary to access the graph for each transaction operation. Oze avoids the giant lock by allowing both a transaction and a record to manage a sub-graph of the MVSG.

Let $T_i$ be a set of transactions that depend on $t_i$ directly or transitively, where for all $t$ in $T_i$, there exists a path from $t_i$ to $t$ in $MVSG(s, \ll)$ through any types of edges, \textit{wr}, \textit{rw}, and \textit{ww}. We refer to $T_i$ as the \textit{followers} of $t_i$. Next, let $SG(s, t_i)$ be the sub-graph of $MVSG(s, \ll)$, consisting of the nodes for $t_i$ and for each transaction in $T_i$ and their edges. We refer to $SG(s, t_i)$ as the \textit{transaction-local} graph of $t_i$. Since $T_i$ includes all transactions reachable from $t_i$, if $SG(s, t_i)$ has no cycle through $t_i$, then $MVSG(s, \ll)$ also has no cycle through $t_i$. Thus, $t_i$ can be safely committed without violating serializability by checking $SG(s, t_i)$.

Let $SG(s, x)$ be the sub-graph of $MVSG(s, \ll)$, consisting of the nodes for each transaction that reads or writes a record $x$ and the \textit{wr}-, \textit{rw}-, and \textit{ww}- dependency edges regarding $x$. We refer to $SG(s, x)$ as the \textit{record-local} graph of $x$. When $X_i$ denotes the set of records read or written by $t_i$ and $T_i$, it follows that \[ SG(s, t_i) = \bigcup_{x \in X_i} SG(s, x). \]

The essence of the Oze protocol is creating $SG(s, t_i)$ by merging a record-local graph $SG(s, x)$ for each $x$ in $X_i$ and checking for acyclicity. We refer to $X_i$ as the \textit{target record set} of $t_i$ for the merging of record-local graphs.

\subsection{Protocol} \label{sec:protocol}

Oze adopts a three-phase protocol, which consists of reading, validating, and finalizing, to smoothly switch to OCC~\cite{Tu13}, which has three similar phases.

\begin{figure}
    \centering
    \includegraphics[keepaspectratio, scale=0.5]{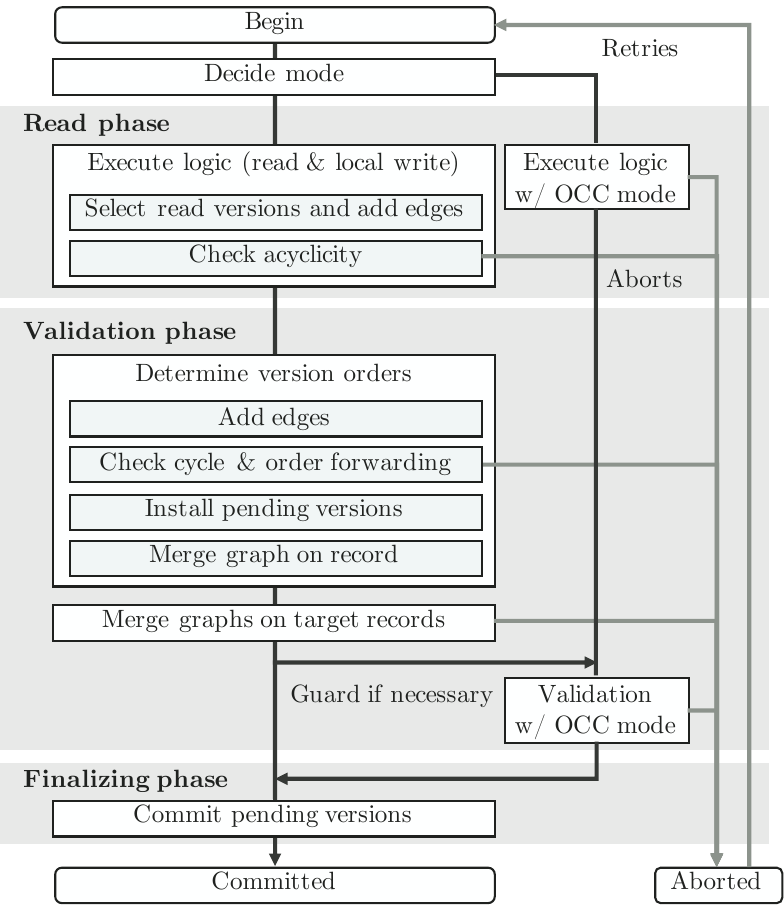}
    \adjustCaption
    \caption{Overview of Oze protocol.}
    \label{fig:protocol}
\end{figure}

\subsubsection{Read phase}
Figure~\ref{fig:protocol} shows an overview of the Oze protocol. Oze executes a transaction while reading committed versions and writing new ones in the thread-local storage. When reading a record $x$, the transaction selects the latest version from the version list sorted by the serialization order, and a \textit{wr}-dependency edge is added to the record-local graph of $x$. If the graph is acyclic (i.e., serializable), the transaction adds the version to the local read set.
If the graph is cyclic, then it finds an older version that does not create a cycle. In such a case, the transaction adds \textit{rw}-dependency edges, which are from the transaction to the writers of the versions newer than the selected one into the record-local graph.

When writing a record $x$, the transaction only adds the new version of the record in the local write set.

\subsubsection{Validation phase} \label{sec:design-validation}
On reaching the commit operation, a transaction $t_i$ enters the validation phase, which consists of two parts: (1) choosing version orders based on the method in~\Cref{sec:version-order} and then (2) merging the record-local graph for each record in $X_i$ to obtain $SG(s, t_i) = \bigcup_{x \in X_i} SG(s, x)$ while identifying followers and checking for acyclicity of the graph. For simplicity, we assume here all concurrent transactions other than $t_i$ are pending during the validation phase of $t_i$. We describe how to handle concurrent transactions without using a global giant lock in~\Cref{sec:read-write,sec:commit}.

(1) Choosing version orders: For each record in the write set, $t_i$ performs the following three steps: (1a) $t_i$ determines a version order and adds the corresponding edges to the record-local graph; (1b) after confirming acyclicity, $t_i$ installs a pending version (which cannot be read by other transactions at this point) into the version list of the record; and (1c) $t_i$ merges the record-local graph into its transaction-local graph.

(2) Merging record-local graphs: If the version order can be determined for all records in the write set while keeping the graph acyclic, $t_i$ then merges the record-local graph for each record in the read set into its transaction-local graph. Each time $t_i$ merges a record-local graph, it checks the acyclicity of the resulting graph and identifies all its followers. If a follower of $t_i$ exists, $t_i$ merges record-local graphs of the records read or written by that follower and repeats identifying its followers and merging the record-local graphs. If no cycle is found in $SG(s, t_i)$, which is obtained by merging all graphs on records read or written by all followers, $t_i$ proceeds to the finalizing phase for committing.

\subsubsection{Finalizing phase}
For a transaction that passed the validation phase without aborting, Oze changes the status of versions written by the transaction to \textit{committed}.

\subsection{Dynamic Version Ordering} \label{sec:version-order}
A notable feature of Oze is dynamic version ordering, which actively explores possible serializable schedules in the validation phase described in~\Cref{sec:design-validation}.

When choosing the version order, Oze first tries postposing the version so that the newer version in chronological order becomes the newer version in the serialization order. Given that there is a transaction $t_i$ that reads $x$ written by transaction $t_j$, postposing $t_k$ means to select the order of $x_j \ll x_k$. If postposing breaks serializability (i.e., a cycle occurs in MVSG), then Oze tries preposing ($x_k \ll x_j$) to find an acyclic schedule. We call this technique order forwarding. With the order forwarding, Oze can change the version order while keeping the concurrent transactions' views, i.e., as long as the forwarding transaction does not interrupt the writer of the version and its readers. The order forwarding enables Oze to schedule transactions in the MVSR space, which is larger than the space generated by the MVSGT protocol~\cite{Hadzilacos85}.

\begin{figure}
    \centering
    \includegraphics[keepaspectratio, scale=0.44]{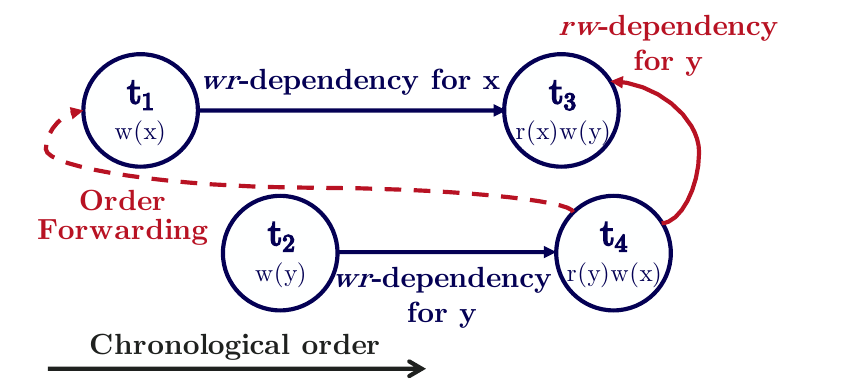}
    \adjustCaption
    \caption{Example of order forwarding.}
    \label{fig:order-forwarding}
\end{figure}

\textbf{Example: }
The following example illustrates the behavior of order forwarding. Consider the schedule $s$, which consists of transactions $t_1$ through $t_4$. Assume that all transactions are concurrent; here, $c$ denotes an internal commit, which occurs before the response is returned to the client. All other notations follow the multi-version full schedule notation in the literature~\cite{Hadzilacos85}.
\begin{equation*}
    s = w_1(x_1) w_2(y_2) c_1 c_2 r_3(x_1) r_4(y_2) w_3(y_3) c_3 w_4(x_4) c_4
\end{equation*}

After committing $t_1$ and $t_2$, $t_3$ and $t_4$ read $x_1$ and $ y_2$, respectively. As a result, the corresponding \textit{wr}-dependency edges (shown as blue edges in Figure~\ref{fig:order-forwarding}) are added. When $t_3$ reaches $c3$, it checks if it can choose $y_2 \ll y_3$ (postposing). This is done by adding an \textit{rw}-dependency edge from $t_4$ to $t_3$ (solid red edge in the figure) and checking for acyclicity. Since no cycle exists in this example, $t_3$ can safely commit.

In contrast, when $t_4$ enters its validation phase at $c_4$ (after $c_3$), $t_4$ creates a cycle when it tries to add a postposing edge from $t_3$ to $t_4$ (i.e., \textit{rw}-dependency edge for the version order $x_1 \ll x_4$) due to $t_3$'s read of $x_1$. Thus, it performs the order forwarding and tries to add a preposing edge from $t_4$ to $t_1$ (i.e., \textit{ww}-dependency edge for the version order $x_4 \ll x_1$) as shown by the dashed red edge in the figure. The order forwarding makes the graph acyclic, thus $t_4$ can also be committed. The final serialization order is $t_2 < t_4 < t_1 < t_3$, which is different from the chronological order and cannot be obtained with MVSGT~\cite{Hadzilacos85}. Note that Oze can still ensure \textit{linearizability}~\cite{Herlihy90} by introducing an epoch and restricting the forwarding within the epoch, as detailed in Section~\ref{sec:choosing-version-order}.

\subsection{Dynamic Protocol Switching} \label{sec:protocol-switching}

Oze using MVSG is specially designed for complex workloads that consist of long-running update transactions and conflicting short transactions. Consequently, using MVSG for workloads without such long transactions is not economical due to its maintenance. In order to practically handle both types of workloads, Oze dynamically switches between the MVSG-based protocol and an OCC protocol, which is based on Silo~\cite{Tu13}, depending on the workload characteristics.
Switching from the OCC mode to MVSG mode starts when a long transaction is aborted. Oze detects the long transaction when its read/write set is over a predefined size. Conversely, switching from the MVSG to OCC mode starts when workers do not observe long transactions for a certain period. Those switchings go through a transition mode so that each worker can continuously process transactions without quiescing. During the transition mode, the workers manage the graphs but validate transactions using the OCC protocol, as shown in the last part of the validation phase in~\Cref{fig:protocol}. This strategy preserves serializability because the OCC mode has a smaller scheduling space than that of the MVSG mode.

\section{Oze Implementation} \label{sec:implementation}
This section describes an implementation of Oze's MVSG mode.

\subsection{Data Structure} \label{sec:data-structure}

\begin{figure}[t]
  \centering
  \includegraphics[keepaspectratio, scale=0.39]{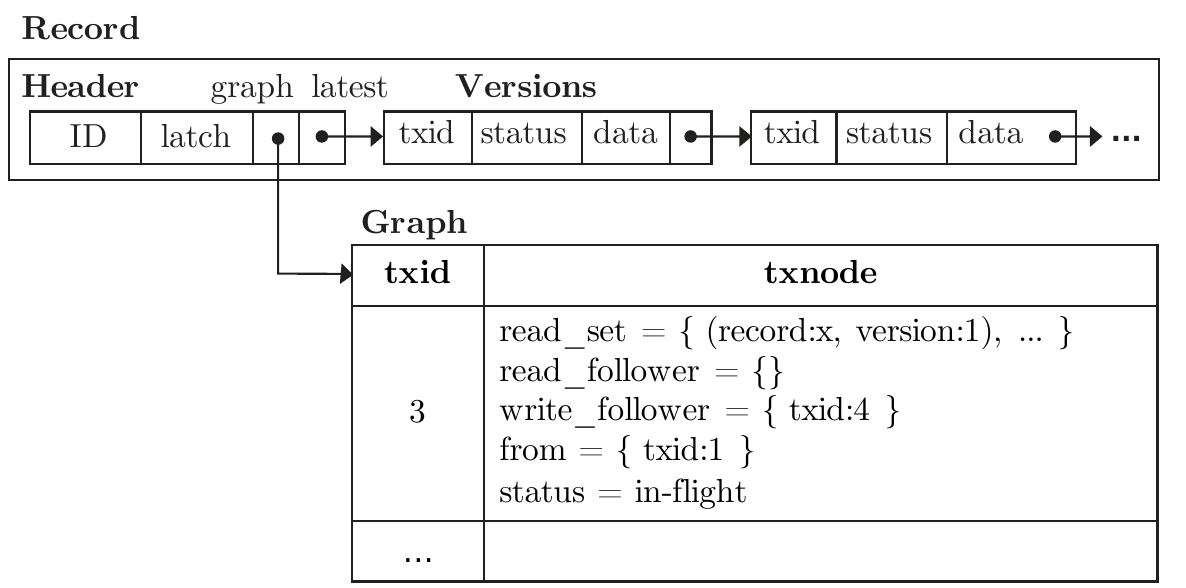}
  \adjustCaption
  \caption{Data structures in Oze.}
  \label{fig:data-structure}
\end{figure}

\textbf{Transaction: }
Each transaction has a transaction ID (\textit{txid}) and read/write sets. The transaction ID is assigned at the beginning, and it consists of an epoch, a worker thread ID, and a local counter. We introduce the epoch to ensure linearizability and facilitate garbage collection. Like Silo~\cite{Tu13}, a dedicated thread increments the global epoch at regular intervals, and each worker thread refers to it. The read/write sets manage entries with a primary key, a record pointer, and a version pointer.

\textbf{Record: }
The structure of a record is shown in Figure~\ref{fig:data-structure}. The record contains a record ID, a latch, and two pointers: one for a linked list of versions and one for a record-local graph.  The record ID is a 64-bit integer used for OCC (i.e., TID in Silo~\cite{Tu13}). The versions in the list are ordered in the serialization order. Each version has \textit{txid} of the version creator, the pointer to the previous version, and the status.

\textbf{Graph: }
The structure of a record-local graph is shown in Figure~\ref{fig:data-structure}, and it is also used for a transaction-local graph. The graph is represented as a map whose key is \textit{txid} and whose value is the node of the graph (\textit{txnode}). Each node has three lists of \textit{txids}. \textit{read\_follower} is a list of transactions that read the version, corresponding to the \textit{wr}-dependency edges. \textit{write\_follower} is a list of transactions that write a version newer than the version read or written by the transaction, corresponding to the \textit{rw}-dependency edges and \textit{ww}-dependency edges. \textit{from} is a list of transactions that point to the list owner, i.e., incoming edges. The node also has the transaction's read set and the status to find followers.

\subsection{Read Phase} \label{sec:read-write}

As described in Section~\ref{sec:protocol}, Oze executes the read phase, validation phase, and finalizing phase in that order. We describe the read and write protocols in the read phase below.

Lines 1--15 of Algorithm~\ref{alg:readwrite} show Oze's read protocol.
This \textit{read()} function is called if a transaction does not have the cached record and version in the local read and write sets. During \textit{read()}, the transaction must hold a latch to access the record exclusively, including the versions and the record-local graph. First, it searches the version list of the record from the latest version to find a \textit{readable version}, which is a committed version that does not create a cycle when reading. If it does not find a readable version even after reaching the oldest version, it aborts. If there is a readable version, it adds its \textit{txid} to the \textit{read\_follower} in the \textit{txnode} of the transaction that wrote the selected version and adds the writer's \textit{txids} of the skipped versions to the \textit{write\_follower} in the own \textit{txnode} (Lines 5--10).
The record and the version are also added to the local read set and the read set on the record-local graph, respectively (Lines 12--13).

When writing, a transaction just creates a version and stores it in the local write set (Lines 17--18).
The versions can be used for \textit{"read-own-write"}~\cite{Orji88,Lim17}.

\begin{algorithm}[t]
  \small
  \caption{Read phase}\label{alg:readwrite}
  \Function{read(txn, record)}{
      latch(record) \\
      graph = record.graph \\
      version = record.latest \\
      \While{version}{
          graph[version.txid].read\_follower.add(txn.txid) \\
          \lIf{is\_acyclic(graph)}{
            break
          }
          graph[version.txid].read\_follower.remove(txn.txid) \\
          graph[txn.txid].write\_follower.add(version.txid) \\
          version = version.next
      }
      \uIf{version}{
          txn.read\_set.add((record, version)) \\
          graph[txn.txid].read\_set.add(record, version) \\
      }
      \lElse{
          abort()
      }
      unlatch(record) \\
  }
  \Function{write(txn, record)}{
      version = create\_version(txn.txid) \\
      txn.write\_set.add((record, version)) \\
  }
\end{algorithm}

\subsection{Validation Phase} \label{sec:commit}

Algorithm~\ref{alg:validation-phase} shows the entire flow of the validation phase. As described in~\Cref{sec:protocol}, it consists of two parts: choosing a version order for each record in the write set (Lines 2--9) and obtaining a union of record-local graphs on records read or written by followers of the transaction in validation (Lines 10--20).

\subsubsection{Choosing version order} \label{sec:choosing-version-order}

For each record in the write set, a transaction first merges its transaction-local graph into the record-local graph to share it with concurrent transactions (Line 4). Sharing the transaction-local graph enables concurrent transactions to abort as early as possible before merging record-local graphs on many records and increases the chance of trying the order forwarding.
After merging, the transaction chooses the version order and checks for acyclicity (Line 5; see also Lines 21--38 for details). Oze uses a per-record latch to access the version list and the record-local graph exclusively (Lines 3--9).

In \textit{choose\_version\_order()}, the transaction gets \textit{readers}, a list of transactions reading the record, based on the graph and adds \textit{rw}-dependency edges; the transaction in validation is placed behind the readers in the serialization order (Lines 23--25). If there is no cycle, it inserts the version as the latest one (Lines 26--27).

If there is a cycle, Oze tries the order forwarding. The transaction removes the current \textit{rw}-de\-pen\-den\-cy edges, which are from \textit{readers} to \textit{txn} itself (Lines 29--30). To try another version order, the transaction finds \textit{writers}, which are transactions that write versions read by \textit{readers} so that \textit{txn} can be ordered before \textit{writers} (Line 31). Specifically, it finds such transactions by checking each \textit{reader's} read set and the writers' transaction ID of the versions in the read sets. Then, the transaction adds new \textit{ww}-dependency edges from \textit{txn} to each of \textit{writers} and checks for acyclicity (Lines 34 and 36). If there is no cycle, the new version is inserted into the version list, considering the order of the other versions written by \textit{writers}.

The order forwarding can be performed across epochs. However, Oze limits it within the same epoch to guarantee linearizability and simplify graph cleaning (described in Section~\ref{sec:gc}) and aborts transactions if it occurs across epochs (Lines 33--35).

For scalability, multiple transactions speculatively choose version orders and write the corresponding edges to each record-local graph in parallel. Thus, Oze's concurrent implementation may cause false-positive aborts, making it more restrictive than the centralized MVSG approach.

If the transaction successfully chooses the version order, it then merges the record-local graph into the transaction-local graph after checking the follower transactions to identify the target records ($X_i$ defined in~\Cref{sec:decentralizing-mvsg}) to be merged later (Lines 6--7). The function \textit{add\_target\_\ records()} in Lines 39--43 illustrates how a transaction identifies the target records. It first retrieves all followers (Line 40) and adds each record in the follower's read set to the target list (Lines 41--43). Note that followers in the read phase can be ignored because they will identify their own followers during their validation phase. It is also unnecessary to inspect the followers' write sets. A follower merges record-local graphs on records in its write set into its transaction-local graph (Line 7) and then leaves it on records in its read set (Line 14). Thus, all \textit{wr}-, \textit{rw}-, and \textit{ww}- dependency edges involving a follower are included in a record-local graph on one of the records in its read set. Therefore, a transaction in validation phase does not need to check the followers' write sets.

\begin{algorithm*}[t]
  \vspace{-1.5em}
  \caption{Validation phase}\label{alg:validation-phase}
  \begin{multicols}{2}
  \small
  \tcp{\textbf{done:} List of records already processed}
  \tcp{\textbf{target:} List of target records $X_i$ for merging graphs}
  \tcp{Both lists are empty at beginning}
  \Function{validate(txn)}{
      \For{(record, version) \textbf{in} write\_set}{
          latch(record) \\
          merge(record.graph, txn.graph) \tcp{record <- txn}
          choose\_version\_order(txn, record, version) \\
          add\_target\_records(txn, record.graph, target, done) \\
          merge(txn.graph, record.graph) \tcp{txn <- record}
          done.add(record) \\
          unlatch(record) \\
      }
      target.add(records in read\_set) \\
      \While{! target.is\_empty()}{
          record = target.pop() \\
          latch(record) \\
          merge(record.graph, txn.graph) \tcp{record <- txn}
          \If{! is\_acyclic(record.graph)}{
              abort()
          }
          add\_target\_records(txn, record.graph, target, done) \\
          merge(txn.graph, record.graph) \tcp{txn <- record}
          done.add(record) \\
          unlatch(record) \\
      }
  }
  \Function{choose\_version\_order(txn, record, version)}{
    graph = record.graph \\
    readers = find\_readers(graph, record) \\
    \For{r \textbf{in} readers}{
        graph[r.txid].write\_follower.add(txn.txid) \\
    }
    \eIf{is\_acyclic(graph)}{
        record.insert\_version(version, []) \\
    }{
        \tcp{Order forwarding}
        \For{r \textbf{in} readers}{
            graph[r.txid].write\_follower.remove(txn.txid) \\
        }
        writers = find\_writers(graph, record, readers) \\
        \For{w \textbf{in} writers}{
            \uIf{txn.txid.epoch == w.txid.epoch}{
                graph[txn.txid].write\_follower.add(w.txid) \\
            }
            \lElse{
                abort()
            }
        }
        \uIf{is\_acyclic(graph)}{
            record.insert\_version(version, writers) \\
        }
        \lElse{
            abort()
        }
    }
  }
  \Function{add\_target\_records(txn, graph, target, done)}{
    followers = get\_followers(txn, graph) \\
    \For{follower \textbf{in} followers}{
        \For{(rec, v) \textbf{in} graph[follower].read\_set}{
            \lIf{rec \textbf{not in} done}{
                target.add(rec)
            }
        }
    }
  }
  \end{multicols}
\end{algorithm*}

\subsubsection{Merging record-local graphs on target records}

Since record-local graphs for each record in the write set are merged when choosing version orders, the transaction then repeatedly merges graphs for each record in the read set while identifying followers and additional target records. Specifically, it merges its transaction-local graph into the record-local graph, checks the acyclicity of the merged graph, lists additional target records to be merged, and continues merging the record-local graph into the transaction-local graph until the target list becomes empty (Lines 11--20).
The transaction leaves its footprints by merging the transaction-local graph into the records for the reason described in~\Cref{sec:choosing-version-order} (see the discussion on Lines 4 and 14 for more details).
Note that the transaction must hold a per-record latch (Lines 13--20).

\subsubsection{Correctness sketch} \label{sec:decentralization-correctness}

If a set of uncommitted transactions $T_{cycle}$ forms a cycle, Oze ensures at least one transaction in the cycle will be aborted. Let $t_i$ in $T_{cycle}$ be a last transaction that determines its version orders and writes the corresponding edges to record-local graphs; i.e., $t_i$ has reached Line 10 of Algorithm~\ref{alg:validation-phase} lastly. Then, at this moment, $t_i$ must exhaustively identify all followers in $T_{cycle}  \subseteq T_i$ and their dependencies, which are already stored in the target record set $X_i$. Therefore, $t_i$ will detect the cycle and abort.

\subsubsection{Complexity} \label{sec:complexity}

The dominant factor of the complexity in the validation phase is graph processing, such as merging and cycle-checking. The time complexities of both processes are $O(|V| + |E|)$ where $|V|$ and $|E|$ are the number of nodes and edges in a graph\footnote{Precisely, the cycle check requires fewer nodes and edges since it is enough to only check the nodes that follow the transaction about to commit.}, respectively. The graph processing must be done for each record in the target record set $X_i$. Therefore, as a whole, the time complexity of the decentralized Oze protocol is $O(|X_i|(|V| + |E|))$.

\subsection{Parallel Validation} \label{sec:parallel-validation}
In Oze, merging record graphs and checking for acyclicity during the validation phase incurs high overhead, particularly for long transactions involving many read and write operations. Moreover, due to infrequent garbage collection during such transactions, the graph size may keep growing, potentially causing the validation process to run indefinitely. To mitigate this, Oze parallelizes the validation phase using multiple threads to accelerate validation.

A part that can be parallelized is merging record-local graphs on the target records (Lines 11--20 in~\Cref{alg:validation-phase}).
Specifically, parallel validation works as follows. (1) A transaction worker thread invokes the specified number of validators and assigns merge targets to validators equally. (2) For each target record, each validator merges the validator-local graph into the record-local graph, checks for acyclicity of the graph, finds additional targets, and merges the graph into the validator-local graph again. (3) Once all validators finish, the transaction worker merges all the validator-local graphs into the transaction-local graph and checks for acyclicity. (4) If the graph remains acyclic, the transaction worker aggregates additional targets identified by each validator and reassigns them equally. Steps (1)--(4) are repeated until no additional targets remain.

\subsection{Phantom Avoidance} \label{sec:phantom-avoidance}

Oze requires phantom avoidance technique that does not depend on an optimistic approach like index node validation~\cite{Tu13} to keep precise tracking of dependencies. We use a variant of precision locking~\cite{Jordan81}, which is also used in Hyper~\cite{Neumann15}. Precision locking preserves predicates of a range scan query and inserted records in the local or global area and prevents phantom anomalies by comparing them. Although in Hyper, scan transactions validate if there are no conflicting insert transactions based on the predicates, in Oze, insert transactions validate scan histories created by other transactions. This opposite way is suitable for Oze because, in Oze, writer transactions decide version order based on potential anti-dependencies, and insert transactions can be handled in the same manner.

Specifically, a transaction stores the \textit{txid} and the query predicates\footnote{In our current implementation, scan queries support range-based predicates for primary keys, but it can be extended for arbitrary predicates on any columns.} when scanning a table as a scan history. An insert transaction (inserter) checks the scan histories in the validation phase. If the inserter finds that its insertion matches the predicates of a scan history, it adds an edge from the transaction that created the scan history to the inserter on the transaction-local graph. Then, it checks for acyclicity and proceeds to validate other records in the write set if no cycle exists.

\subsection{Garbage Collection} \label{sec:gc}

In SGT~\cite{Casanova80}, once a transaction commits, no incoming edges are added to the node of the transaction. This property allows safe garbage collection of committed nodes. Similarly, we can delete MVSG nodes that will never make a future cycle if we ensure that no new incoming edges will be added. However, in Oze, incoming edges can be added to the committed transaction in two situations. The first case can occur in the read protocol. As mentioned in Section~\ref{sec:read-write}, when selecting a version to read, a reader transaction may add edges to the transactions (\textit{write\_follower}) that wrote the skipped version.
The second case can occur in the order forwarding; when preposing a transaction, it adds edges to the \textit{write\_follower} transactions as mentioned in Section~\ref{sec:commit}.

Oze uses epochs to guarantee that no new incoming edges will be added to a node, allowing it to be safely deleted. An epoch that satisfies this condition is referred to as the \textit{reclamation epoch} ($e_r$), defined as the minimum local epoch among all worker threads minus one. To enforce this, Oze prevents transactions from adding incoming edges by prohibiting: (1) reading versions from $e_r$ or earlier (except for the latest version), and
(2) performing the order forwarding to versions in $e_r$ or earlier.

For GC of versions, we remove versions in $e_r$ or earlier, except for the latest (in the serialization order). We must check the MVSG and hold versions if concurrent transactions are reading them.

\section{Evaluation} \label{sec:eval}

This section shows (1) Oze can handle long-running update transactions while achieving better short-transaction performance and smoothly switching from/to OCC in our target workload, BoMB, (2) Oze achieves comparable performance to existing protocols even in TPC-C~\cite{tpcc}, (3) Oze's order forwarding works effectively in a specific workload, and (4) Oze has the expected computational complexity through a runtime analysis with YCSB~\cite{Cooper10}.

\subsection{Experimental Setup}

All experiments were performed using CCBench~\cite{Tanabe20}, a benchmark platform for various concurrency control protocols.
We implemented two variations of Oze in CCBench: one used the proposed decentralized MVSG, and the other one used a single Centralized MVSG (Oze-CM).
We compared Oze with three OCC protocols: Silo~\cite{Tu13}, MOCC~\cite{Wang16}, TicToc~\cite{Yu16} and two MVCC protocols: ERMIA~\cite{Kim16} as SSN and Cicada~\cite{Lim17}. We also used two pessimistic approaches: 2PL-No-Wait~\cite{Bernstein81} and D2PL. D2PL is a 2PL-based protocol that mimics deterministic behavior such as in Calvin~\cite{Thomson12}. D2PL first sorts all the accessing keys and then locks them in that order to avoid deadlocks.
Each protocol had the same interfaces, such as begin, commit, read (point queries), scan (range queries), etc, and we implemented various transactions in workload programs using those interfaces. All protocols incorporated the adaptive backoff mechanism~\cite{Lim17,Tanabe20} to reduce contention.

The evaluation environment consisted of a single server with two Intel\textregistered Xeon\textregistered Gold 6138 CPUs with 2.00 GHz and twenty-four DDR4-2666 32 GB DIMMs (total 768 GB). Each CPU had 20 cores.

\begin{figure*}
  \begin{minipage}{0.32\linewidth}
    \centering
    \includegraphics[keepaspectratio, scale=0.47]{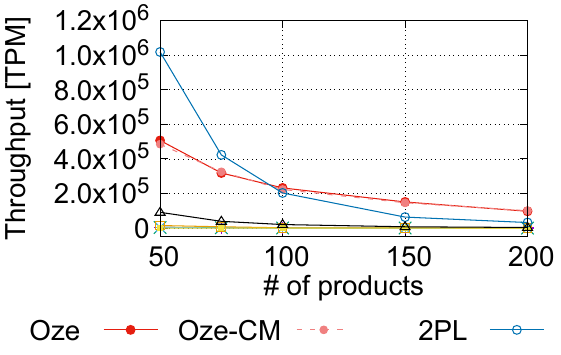}
    \subcaption{L1 length scalability with L1 abort rate $<1$\% (1 thread for L1 and 1 thread for other short transactions)}
  \end{minipage}
  \hspace{0.015\columnwidth}
  \begin{minipage}{0.32\linewidth}
    \centering
    \includegraphics[keepaspectratio, scale=0.47]{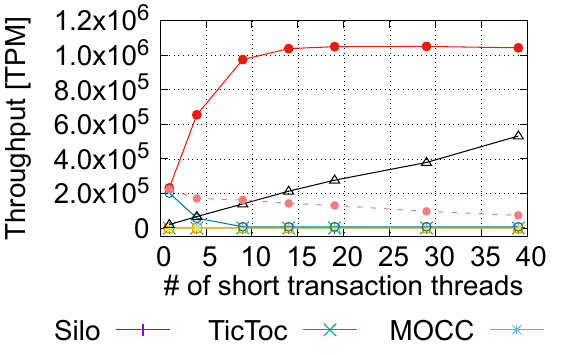}
    \subcaption{Thread scalability with 100 products and L1 abort rate $<1$\% (1 thread for L1 and varying \# threads for short transactions)}
  \end{minipage}
  \hspace{0.015\columnwidth}
  \begin{minipage}{0.32\linewidth}
    \centering
    \includegraphics[keepaspectratio, scale=0.47]{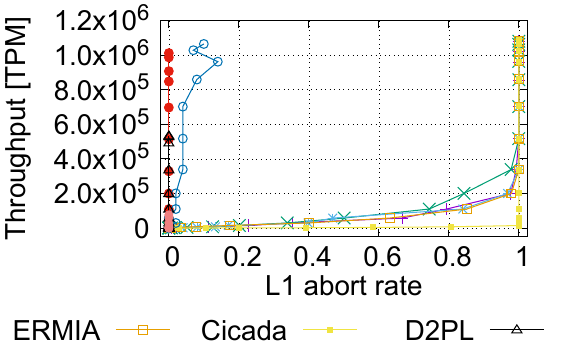}
    \subcaption{TPM vs. L1 abort rate with 100 products, 40 threads (1 thread for L1 and others for short transactions), and varying request rate}
  \end{minipage}
  \adjustCaption
  \caption{Short transaction throughput with static BoM.}
  \label{fig:sbom-cc-comparison}
\end{figure*}
  
\begin{figure*}
  \begin{minipage}{0.32\linewidth}
    \centering
    \includegraphics[keepaspectratio, scale=0.47]{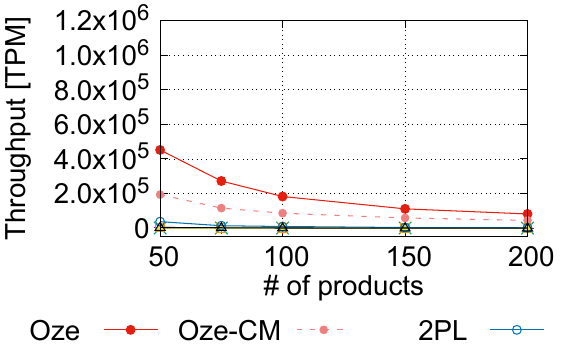}
    \subcaption{L1 length scalability with L1 abort rate $<1$\% (1 thread for L1 and 1 thread for other short transactions)}
  \end{minipage}
  \hspace{0.015\columnwidth}
  \begin{minipage}{0.32\linewidth}
    \centering
    \includegraphics[keepaspectratio, scale=0.47]{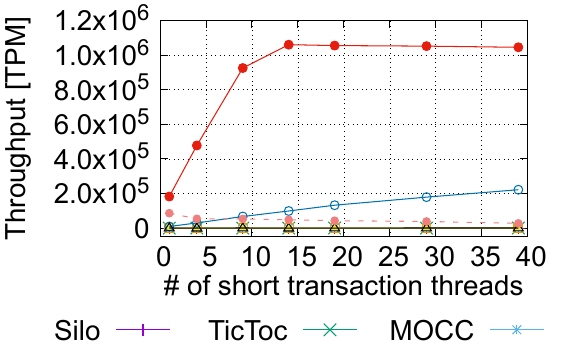}
    \subcaption{Thread scalability with 100 products and L1 abort rate $<1$\% (1 thread for L1 and varying \# threads for short transactions)}
  \end{minipage}
  \hspace{0.015\columnwidth}
  \begin{minipage}{0.32\linewidth}
    \centering
    \includegraphics[keepaspectratio, scale=0.47]{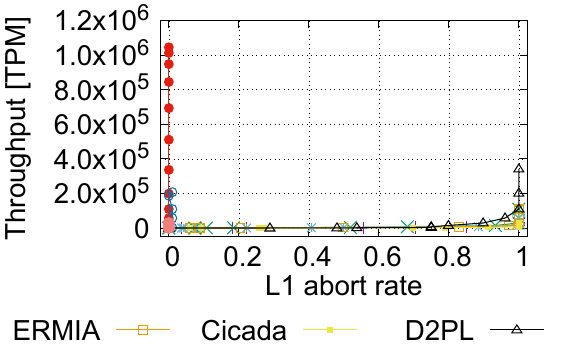}
    \subcaption{TPM vs. L1 abort rate with 100 products, 40 threads (1 thread for L1 and others for short transactions), and varying request rate}
  \end{minipage}
  \adjustCaption
  \caption{Short transaction throughput with dynamic BoM.}
  \label{fig:dbom-cc-comparison}
\end{figure*}

\subsection{Experiments on BoMB} \label{sec:eval-bomb}

We first evaluated how each protocol could handle the BoMB workload using \textit{static BoM}, i.e., L1, S1, and S2 transactions only; BoM trees never changed. Then, we ran all six transactions of \textit{dynamic BoM}, including transactions that changed BoM trees.

\subsubsection{Static BoM} \label{sec:sbom-results}

We measured the maximum short transaction throughput based on the regulation we defined in~\Cref{sec:bomb-transaction}. We used the default parameters of the BoMB~\cite{BoMB}. We increased the request rate of short transactions (with a 50/50 ratio for S1/S2) as much as possible, keeping the L1 abort rate less than 1\% based on the requirement. We stopped increasing the request rate when the average throughput did not change by more than 5\% and used the effective throughput at that time as the score. Starting with one transaction per second, we increased the request rate by a factor of 2 and ran the execution for 60 seconds.
We used the average throughput of three runs.
Note that L1 transactions were always running during each execution on one thread.
For Oze, we used a 40ms epoch length in all experiments because our preliminary experiments did not show significant differences between 40ms and 5ms. We also used 32 threads for parallel validation.

\textbf{L1 length scalability:} \Cref{fig:sbom-cc-comparison}(a) shows the average throughput of short transactions (S1 and S2) while increasing the number of target products from 50 to 200. The number of worker threads was one for the L1 transactions and one for the short transactions. OCC protocols (i.e., Silo, TicToc, and MOCC) performed worse than others by several orders of magnitude since the L1 validation phase inevitably failed when increasing the S1 requests, which updated the cost of many raw materials. MVCC protocols (i.e., ERMIA and Cicada) also performed worse. L1 could proceed to build BoMs and calculate costs without being hindered by S1, thanks to the benefit of multiple versions. However, L1 frequently aborted with false positives when L1 tried to update the costing results that conflicted with the S2's reads. 2PL and D2PL worked better than OCC and MVCC when the L1 length was shorter; however, they reduced throughput as L1 became longer since short transactions that conflicted with L1 had to wait to acquire locks. When increasing products, Oze showed the highest throughput.
Oze with a single centralized MVSG (Oze-CM) showed slightly lower throughput than decentralized Oze because the single-threaded short transaction did not cause much contention on the graph.

\textbf{Thread scalability:} \Cref{fig:sbom-cc-comparison}(b) shows the short transaction throughput with the fixed product size of 100 for the L1 transaction while increasing the number of worker threads for short transactions. The other measurement process was the same as the above. The protocols other than lock-based approaches and Oze could not benefit from parallelism since a single thread was enough to hinder L1. The throughput of Oze-CM gradually decreased due to contention on the graph as the number of worker threads increased. 2PL also reduced the throughput because the abort rate of the L1 transaction exceeded the requirement of 1\% when increasing the threads. Oze and D2PL, where L1 did not abort at all, increased the short transaction throughput depending on the number of threads. D2PL was still inferior to Oze because short transactions with D2PL required a long time to acquire locks due to L1.

\textbf{L1 abort rate:} In \Cref{fig:sbom-cc-comparison}(c), we plotted the L1 abort rate (x-axis) against the short transaction throughput (y-axis). We fixed the number of products to 100 and the number of threads at 40, then increased the request rate until the throughput reached saturation, ignoring the 1\% L1 abort rate regulation in BoMB. Although OCC and MVCC protocols increased throughput at the right end of the graph, the throughput reached saturation as the same as Oze. In contrast, Oze increased the throughput while committing the L1 perfectly, as all plots are on the left end.

\textbf{Key finding with static BoM:} The state-of-the-art optimistic protocols, including MVCC, hardly commit long-running update transactions in the BoM workload even if the BoM tree does not change. Oze and D2PL handle the static BoM workload well. 2PL can commit the L1 transactions in some cases.

\subsubsection{Dynamic BoM} \label{sec:dbom-results}

Using the same procedure as with the static BoM, we measured the short transaction throughput with the ratio defined in~\Cref{sec:bomb-transaction}.
For the dynamic BoM, we used different phantom protection techniques for each protocol. For 2PL, we used a variant of gap-locking~\cite{Lomet93}. For D2PL, we issued a reconnaissance query~\cite{Thomson12} and validated the results. For Oze, we used a variant of precision locking described in~\Cref{sec:phantom-avoidance}. For other protocols, we used index node validation~\cite{Tu13}. Figure~\ref{fig:dbom-cc-comparison} shows the results of the same three experiments in~\Cref{sec:sbom-results} in the dynamic BoM setting. Note that ~\Cref{fig:cc-comparison-tps-vs-latency} plots the throughput and latency using the results of \Cref{fig:dbom-cc-comparison}(c) as long as the L1 abort rate is less than 1\%.

\textbf{L1 length scalability:} OCC and MVCC protocols performed worse, similar to the case in the static BoM. D2PL also could not handle the BoMB workload because the validation of the reconnaissance query almost always failed due to the S3 and S4 transactions. 2PL significantly reduced throughput due to the wait for the locks held by the L1 transaction since the S3 and S5 try to update the target products of the cost calculation. In contrast, Oze could handle L1 the same as in the static BoM.

\textbf{Thread scalability:} For protocols other than D2PL and 2PL, we observed similar results to the ones in the static BoM. For D2PL, additional threads no longer affected the throughput of D2PL due to the same reason in the single-thread case of Figure~\ref{fig:dbom-cc-comparison}(a). Surprisingly, 2PL increased the throughput while committing almost all of the L1 transactions because S3 and S5 transactions reduce the chances of the S2 execution, which hinders L1. Nevertheless, Oze still achieved about five times higher throughput than 2PL.

\textbf{L1 abort rate:} As shown in the figure, existing OCC and MVCC protocols performed worse than those in the static BoM. Thus, we terminated increasing the request rate when the average L1 abort rate continued to be 100\% three times for those protocols. For 2PL, as discussed in the thread scalability above, we saw a significant reduction of the L1 abort.

\textbf{Key finding with dynamic BoM:} Except for Oze, only 2PL can handle the dynamic BoM workload without aborting almost all L1 transactions. However, the short transaction throughput of 2PL is significantly lower than that of Oze because S3 and S5 transactions are stalled for a long time due to a lock held by L1.

\begin{figure}
    \centering
    \includegraphics[keepaspectratio, scale=0.42]{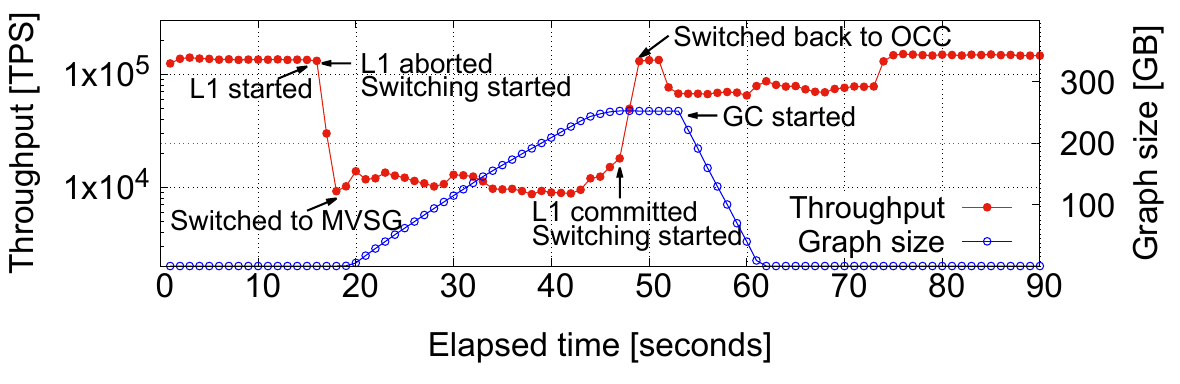}
    \adjustCaption
    \caption{Protocol switching and memory consumption.}
    \label{fig:protocol-switch}
\end{figure}

\subsubsection{Protocol switching} \label{sec:eval-switching}

Figure~\ref{fig:protocol-switch} shows the behavior of protocol switching and memory consumption. The x-axis shows the elapsed time, and the y-axis shows the short transaction throughput (left) and the total size of MVSG graphs for all records (right).

We started ten threads with the OCC mode for only short transactions with a 50/50 ratio for S1/S2. We then submitted an L1 transaction after 15 seconds. The L1 transaction was aborted in the validation phase of the OCC mode about 1 second later, and the threads started switching to the MVSG mode. We confirmed that all the threads smoothly changed to the MVSG mode after a few seconds while reducing throughput, and the L1 was successfully committed after 30 seconds. While executing the L1, it spent almost all the time on the validation phase, and the graph size linearly increased due to merging the graphs on the target record set.
Finally, all the threads switched back to the OCC mode a few seconds after the L1 commit because we configured the threads to switch to the OCC mode when no long transaction exists in a single epoch.

Once all the threads switch to the OCC mode, garbage collection for the graphs starts. We used 32 threads for garbage collection, and it was completed within less than 10 seconds. Note that the short transaction throughput decreased during garbage collection and page reclamation by the operating system.

\subsubsection{Parallel Validation}

\begin{figure}[t]
    \begin{minipage}{0.48\columnwidth}
      \centering
      \includegraphics[keepaspectratio, scale=0.43]{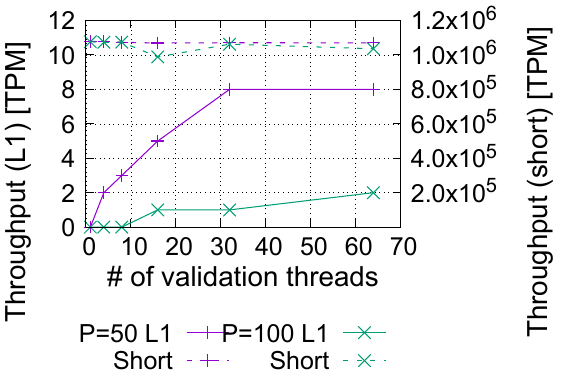}
      \caption{Validation scalability of Oze with static BoM.}
      \label{fig:oze-L1-vscale}
    \end{minipage}
    \adjustCaption
    \hspace{0.02\columnwidth}
    \begin{minipage}{0.48\columnwidth}
      \centering
      \includegraphics[keepaspectratio, scale=0.342]{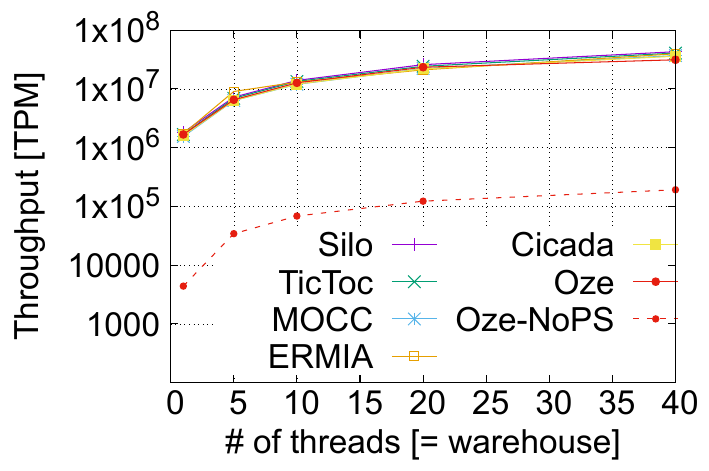}
      \caption{TPC-C throughput (full mix).}
      \label{fig:tpcc}
    \end{minipage}
    \adjustCaption
\end{figure}

Figure~\ref{fig:oze-L1-vscale} shows the throughput of the L1 transaction with 50 and 100 products (P=50 and P=100, respectively) in the static BoM setting using 20 worker threads, as the number of validator threads increases from 1 to 64. The left y-axis shows the throughput for the L1 transaction, and the right y-axis shows the throughput for short transactions.

Oze could not commit the L1 transaction with a single validation thread due to timeouts (i.e., not abort). However, parallel validation enabled L1 commits, improving throughput with more threads. The benefit was reduced as the number of threads increased because the parallel validation presented an overhead when merging each validator's resulting graph and checking for acyclicity.

For short transactions, throughput remained stable even as the number of validator threads increased. This was due to the saturation of short transaction throughput at around ten threads, as shown in \Cref{fig:sbom-cc-comparison}(b), allowing L1 to utilize the remaining resources.

\subsection{Experiments on TPC-C} \label{sec:eval-tpcc}

We compared Oze with other protocols using the TPC-C benchmark, executing all five transactions in its specified ratio. Figure~\ref{fig:tpcc} shows the throughput of each protocol with varying numbers of threads and warehouses. Note that Oze started with the OCC mode, and Oze-NoPS disabled protocol switching. As shown in the figure, Oze-NoPS exhibited poor performance due to its heavy graph management.
Oze showed reasonable performance with existing protocols thanks to protocol switching, although the peak throughput was about 27\% lower than the highest one, Silo.

\subsection{Experiments for Order Forwarding}

In the experiments with BoMB and TPC-C, we could not observe the effects of the order forwarding because there is no such read and blind write combination that triggers it. To confirm the effects, we conducted experiments using synthetic workloads that included four types of transactions, as exemplified in Section 3.3. The target data was a table with 25, 50, or 100 records, each with a key of consecutive integers. T1 and T2 wrote one of the first and last half records, respectively. T3 read one of the first-half records and wrote one of the last-half records, and T4 vice versa. Each transaction type was executed equally. We randomly inserted 0-5ms, 0-10ms, or 0-20ms delays (sleep) after each operation.

\begin{figure}[t]
  \begin{minipage}{0.49\linewidth}
    \centering
    \includegraphics[keepaspectratio, scale=0.345]{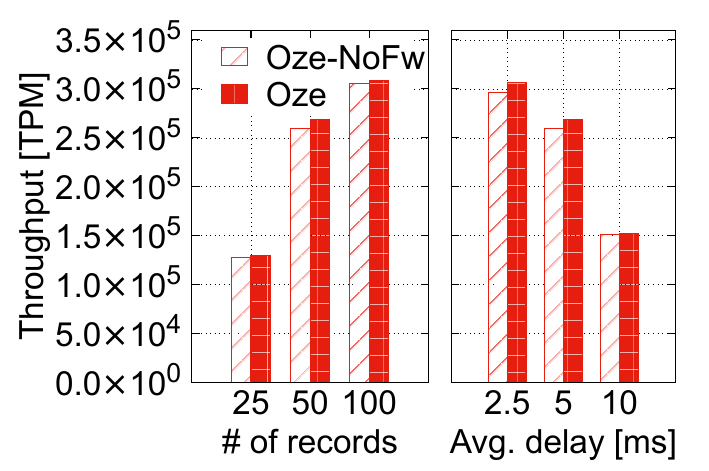}
    \subcaption{Throughput.}
  \end{minipage}
  \begin{minipage}{0.49\linewidth}
    \centering
    \includegraphics[keepaspectratio, scale=0.345]{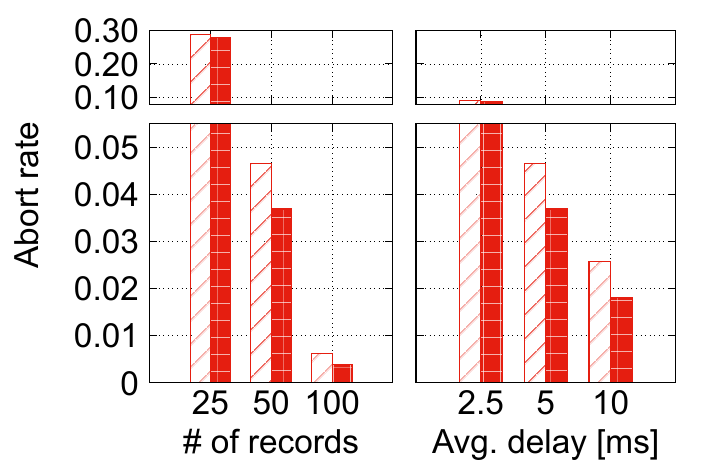}
    \subcaption{Abort rate}
  \end{minipage}
  \adjustCaption
  \caption{Effects of order forwarding.}
  \label{fig:ofwb}
\end{figure}

\Cref{fig:ofwb}(a)(b) show Oze's throughput and abort rate when running the workload with 40 threads in MVSG mode. In Oze-NoFw, the order forwarding was disabled. The left side of each graph shows results under a 5ms average delay while varying the number of records; the right side shows results with 50 records under varying delays. The order forwarding improved throughput by about 3\% and reduced the abort rate by about 20\%. The effect was negligible when the number of records was small (e.g., fewer than the number of threads), due to increased conflicts between transactions of the same type. Improvements were also limited with many records, as transactions committed without requiring the order forwarding.

\subsection{Runtime Analysis with YCSB}

Finally, we conducted an experiment using a YCSB variant for Oze protocol analysis.
Like existing work~\cite{Wang17,Lim17,Tanabe20}, we modified YCSB-A so that each transaction included 50\% reads and 50\% pure writes and ran it on 100 million records with a uniform request distribution.
We intentionally used small payloads (4 bytes) to observe pure overheads of concurrency control. The number of worker threads was 20, and all workers used the single-thread validation.

\begin{figure}[t]
  \begin{minipage}{0.49\linewidth}
    \centering
    \includegraphics[keepaspectratio, scale=0.46]{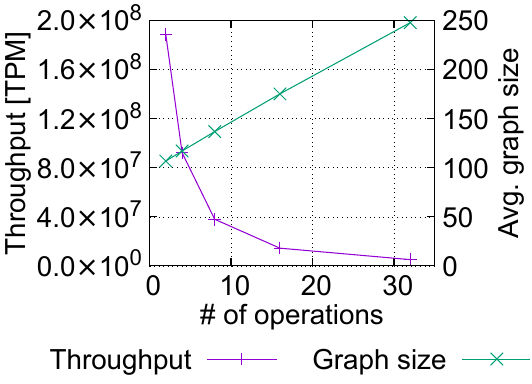}
    \subcaption{Throughput and graph size.}
  \end{minipage}
  \begin{minipage}{0.49\linewidth}
    \centering
    \includegraphics[keepaspectratio, scale=0.46]{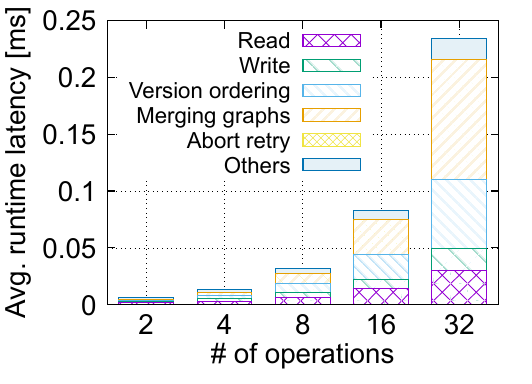}
    \subcaption{Runtime analysis.}
  \end{minipage}
  \adjustCaption
  \caption{Protocol analysis of Oze with YCSB-A.}
  \label{fig:ycsb}
\end{figure}

Figure~\ref{fig:ycsb}(a) shows the throughput and the average graph size as the number of operations per transaction varies. The graph size is the average number of nodes involved in cycle checking. As shown in the analysis of computational complexity in~\Cref{sec:complexity}, the throughput was nearly inversely proportional to the product of the number of operations and the graph size. Figure~\ref{fig:ycsb}(b) shows the average latency required to process each major function of Oze. The percentage of time spent on merging orders became larger compared to other functions as the graph size grew.

\section{Related Work} \label{sec:related}
\textbf{Benchmark: }
TPC-C~\cite{tpcc} and TPC-E~\cite{tpce} are benchmarks for OLTP systems with only short transactions. TPC-EH~\cite{Wang17} has a read-mostly long transaction with write operations, but it lacks a transaction that reads the results written by the long transaction, unlike BoMB. YCSB~\cite{Cooper10}, OLTP-Bench~\cite{Difallah13}, and OLxPBench~\cite{Kang23} provide synthetic and real-world workloads for benchmarking cloud services, OLTP, and HTAP systems, but none of them emulate BoMB's target characteristics. BoMB provides a long-running update transaction that uniquely stresses serializable protocols due to two consecutive and long-term anti-dependency edges created by the other five short transactions.

\textbf{Lock-based protocols: }
2PL variants using timestamp-based priority \cite{Rosenkrantz78,Corbett13,Guo21} can commit a long transaction like L1 when the long transaction gets the smallest timestamp. However, subsequent short transactions must either wait or abort until the long ones commit if they conflict. Altruistic locking~\cite{Salem94} allows transactions to donate locked objects, which other transactions can access before they are unlocked. However, this can increase waiting times for short transactions that accept donations.

\textbf{Graph-based protocols: }
Durner and Neumann proposed a graph-based concurrency control~\cite{Durner19} for many-core systems using Serialization Graph Testing (SGT)~\cite{Weikum01}. Although they extended it as a multi-version concurrency control for read-only analytical queries, their SGT and the multi-version extension cannot handle long-running update transactions and conflicting short transactions concurrently because they always keep the commit order as the constraint order, which narrows the scheduling space.

\textbf{Protocols for highly-contended workloads: }
Protocols like ERMIA~\cite{Yu14} and Cicada~\cite{Lim17} keep multiple versions to handle highly-contended workloads. Hybrid protocols like MOCC~\cite{Wang16} and ACC~\cite{Tang17} switch optimistic and pessimistic schemes to avoid starvation. Techniques like commit-time updates~\cite{Huang20} and batching and reordering schemes~\cite{Ding18} have also been used to avoid high contention. However, there is less discussion on heterogeneous workloads with long-running update transactions, such as BoMB.

\textbf{Lower isolation level protocols with robustness: }
Many \textit{robustness} approaches, which guarantee serializability with a lower isolation level for specific given workloads, have been proposed~\cite{Gan20,Vandevoort21}. Robustness approaches for the case of mixing multiple isolation levels like \textit{serializable} and \textit{snapshot isolation} have also been explored~\cite{Fekete05robustness,Vandevoort23}. We focus on a general-purpose technique guaranteeing serializability for any workloads, even if they are unknown.

\textbf{Deterministic database systems: }
Calvin~\cite{Thomson12} executes transactions based on a pre-determined total order.
Ocean Vista~\cite{Fan19} and Aria~\cite{Lu20} offer alternative approaches that do not require the read-set in advance. When using them for BoMB, however, L1 transactions deteriorate short transaction throughput as they must wait for resolving functors (Ocean Vista) or committing the previous batch (Aria). Oze is a non-deterministic protocol, and its scheduling space is wider than the deterministic ones~\cite{Thomson12,Fan19,Lu20}.

\section{Conclusion} \label{sec:conclusion}

We proposed Oze, a concurrency control protocol that exploits a large scheduling space using a multi-version serialization graph in a decentralized manner. We also proposed an OLTP benchmark, BoMB, based on a use case in an actual manufacturing company. Experiments using BoMB showed that Oze could handle the long-running update transaction while achieving four orders of magnitude higher throughput than optimistic and multi-version protocols and up to five times higher throughput than pessimistic protocols. BoMB and all benchmarked protocols are available on GitHub~\cite{BoMB}.

\begin{acks}
This paper is based on results obtained from the project "Research and Development Project of the Enhanced Infrastructures for Post-5G Information and Communication Systems (JPNP20017)" and JPNP16007 commissioned by the New Energy and Industrial Technology Development Organization (NEDO), and from JSPS KAKENHI Grant Number 25H00446, and from JST CREST Grant Number JPMJCR24R4 and from SECOM Science and Technology Foundation and JST COI-NEXT SQAI (JPMJPF2221), JST Moonshot R\&D Grant Number JPMJMS2215.
\end{acks}


\balance
\bibliographystyle{ACM-Reference-Format}
\bibliography{bibliography}

\end{document}